\documentclass[aps,pra,twocolumn]{revtex4-1}
\usepackage{graphicx}
\usepackage{amsmath}
\usepackage{amsfonts}
\usepackage{amssymb}
\usepackage{mathrsfs}
\usepackage{color}
\usepackage{dsfont} 
\newcommand{\ket}[1]{|#1\rangle}
\newcommand{\bra}[1]{\langle #1|}

\begin{document}
\title{Geometric properties of concatenated geometric 
and holonomic quantum gates}
\author{Rasmus Wallentin}
\author{Erik Sj\"oqvist}
\affiliation{Department of Physics and Astronomy, 
Uppsala University, Box 524, Se-751 20 Uppsala, Sweden}
\begin{abstract}
We investigate the geometric properties of concatenated 
quantum gates, focusing on the distinction between 
geometric and holonomic quantum computation. We show 
that, although individual gates may be purely geometric, 
their concatenation does not generally preserve this 
property, since the cyclic states of the composite 
operation can accumulate nontrivial dynamical phases 
over the course of the concatenated evolution. In 
contrast, we demonstrate that concatenations of 
arbitrary single-qubit holonomic gates remain holonomic, 
provided that the parallel-transport condition holds 
throughout the evolution and the computational subspace 
undergoes cyclic evolution. The preservation of 
geometric or holonomic character under concatenation is 
an important prerequisite for extending arguments 
underlying geometric protection from individual gates to 
complete quantum circuits. 
\end{abstract}
\today

\maketitle

\section{Introduction} 
A quantum gate is said to be geometric or holonomic 
when its action is determined by geometric properties 
of the evolution path in the appropriate state space 
\cite{zhang23}. Quantum circuits consisting of 
geometric gates implemented under Abelian conditions 
\cite{berry84,aharonov87} constitute geometric quantum 
computation (GQC) \cite{ekert00,wang01}. Under 
non-Abelian conditions \cite{wilczek84,anandan88}, 
such circuits give rise to holonomic quantum 
computation (HQC) \cite{zanardi99,sjoqvist12}. 
These two approaches are distinguished by the 
underlying projective-geometric properties of quantum 
states and subspaces, respectively \cite{kult06}. This 
distinction does not imply that the two approaches 
necessarily generate different unitary transformations: 
identical quantum gates can arise from Abelian geometric 
and non-Abelian holonomic implementations.

Geometric and holonomic quantum gates can be 
implemented by identifying evolution paths along 
which the dynamical contribution vanishes, e.g., 
by choosing geodesic segments on Bloch sphere 
\cite{tian04,ota09a,thomas11,zhao21}, Schmidt sphere 
\cite{loredo14,saarijarvi24}, or Grassmann manifold 
\cite{mousolou14,herterich16,xu18}. An alternative 
approach, particularly suited to adiabatic geometric 
quantum computation \cite{ekert00}, employs refocusing 
techniques. In this approach, two loops are implemented 
so that their combined dynamical phases factor out, 
leaving only geometric phases in the resulting state 
transformation \cite{jones00,falci00,azimi16}. A third 
approach involves carefully tuning the Hamiltonian 
parameters and evolution time so that the accumulated 
dynamical phases become trivial upon completion of 
the loops in state space \cite{zhu03,ota09b}. Because 
geometric gates are determined by paths in state 
space, they are expected to be robust against 
certain fluctuations that do not significantly 
alter these paths 
\cite{zhu05,johansson12,kleissler18,alves22}. The extent 
of this robustness in physical implementations of 
geometric and holonomic schemes remains a subject of 
debate \cite{zheng16,colmenar22}. Nevertheless, 
geometric and holonomic gates may prove useful when 
combined with other error-resilient techniques 
\cite{wu05,xu12,zhang14,xu14,wu20,chen21}.

An important consideration in this context is 
whether the geometric character of individual gates 
is preserved when they are combined into a quantum 
circuit. If concatenation introduces non-trivial 
dynamical contributions, the resulting operation is 
no longer purely geometric, and the geometric 
arguments underlying its potential robustness 
may no longer apply to the complete operation. 
Conversely, preserving geometric or holonomic 
character throughout a circuit provides a consistent 
framework for exploiting geometric properties in 
combination with other error-resilient techniques. 

To address the issue of gate concatenation, we 
adopt a global approach to non-adiabatic geometric 
and holonomic quantum computation, extending the 
conventional geometric perspective from individual 
gates to sequences of concatenated gates. We first 
investigate sequences of gates based on Abelian 
geometric phases. In general, the successive 
application of two such gates produces a combined 
operation whose cyclic states differ from those of 
the individual gates. For the combined operation to 
be geometric, these new cyclic states must evolve so 
that their dynamical phases become trivial. Secondly, 
in holonomic quantum computation, by contrast, the 
relevant criterion is determined by the evolution 
of the cyclically evolving computational subspace. 
In the non-adiabatic, non-Abelian setting considered 
here, refocusing and parameter-tuning techniques 
cannot, in general, eliminate the dynamical contribution 
while preserving the holonomic character of the 
evolution. Instead, the dynamical contribution must 
either be trivial or remain proportional to the 
holonomic part throughout the entire evolution 
\cite{fredriksson26a}. In either case, the holonomic 
character of the operation is independent of the 
choice of eigenbasis of the resulting unitary 
operator and is instead determined by the cyclic 
evolution of the computational subspace.

This distinction is also relevant from the perspective 
of robustness. In geometric quantum computation, the 
emergence of nontrivial dynamical phases upon 
concatenation can compromise the purely geometric 
character of the complete operation, potentially 
undermining the geometric basis for its robustness. 
In holonomic quantum computation, by contrast, 
preserving the parallel-transport condition and 
cyclic evolution of the computational subspace 
ensures that the complete operation remains 
holonomic, even when successive gates involve 
different computational bases. Establishing these 
properties at the level of concatenated operations 
is therefore important for assessing the potential 
robustness of geometric quantum circuits and their 
compatibility with additional error-resilient 
techniques.

The paper is organized as follows. In 
Sec.~\ref{sec:Definitions}, we introduce the definitions 
of geometric and holonomic quantum gates and establish 
the corresponding criteria for their implementation. 
In Sec.~\ref{sec:Successive combinations}, we 
investigate the concatenation of geometric and 
holonomic gates and use explicit single-qubit 
examples to illustrate the fundamental distinction 
between the two approaches. Finally, in 
Sec.~\ref{sec:Conclusions}, we summarize our results 
and discuss their implications for geometric and 
holonomic quantum computation.

\section{Geometric and holonomic quantum computation}
\label{sec:Definitions}

\subsection{Geometric quantum gates}
\label{sec:def_GQC}
We consider a quantum system with Hilbert space 
$\mathscr{H}$, whose time evolution is governed 
by the Schr\"odinger equation 
\begin{eqnarray}
\label{eq:S.Eq.}
i\hbar \dot{U} (t,0) = H(t) U(t,0) 
\end{eqnarray}  
with the initial condition $U(0,0)=\hat{1}$, where 
$H(t)$ is a time-dependent Hermitian Hamiltonian. 
The formal solution is given by $U(t,0) = 
\mathcal{T} e^{-(i/\hbar)\int_0^t H(t') dt'}$, 
where $\mathcal{T}$ denotes time-ordering. At 
the final time $T$, the evolution operator 
$U(T,0)$ satisfies the eigenvalue equation 
\begin{eqnarray}
U(T,0) \ket{\psi_k} = e^{i\xi_k} \ket{\psi_k}, 
\ k=1,\ldots,N. 
\label{eq:eigenvalue}
\end{eqnarray}
Here, $N$ denotes the dimension of the subspace 
$\nu_0$, which is invariant under the time 
evolution generated by $U(t,0)$. More precisely, 
the associated projector $P_0$ satisfies 
\begin{eqnarray}
U(t,0) P_0 U^{\dagger} (t,0) = P_0, 
\quad \forall t\in [0,T] .
\end{eqnarray}
Since $U(T,0)$ is a normal operator, its eigenvectors 
can be chosen to form an orthonormal basis of 
$\nu_0$ \cite{nielsen00}. We refer to $\nu_0$ as 
the computational subspace in the following.

For each $\ket{\psi_k(t)}=U(t,0)\ket{\psi_k}$, let 
\begin{eqnarray}
\ket{\psi_k(t)}=e^{if_k (t)} \ket{\varphi_k(t)}
\label{eq:abeliansection}
\end{eqnarray}
such that $f_k(T) = f_k(0) + \xi_k$. Using 
Eqs.~\eqref{eq:eigenvalue} and
\eqref{eq:abeliansection}, we obtain
\begin{eqnarray}
\ket{\varphi_k (T)} = e^{i[-f_k(T) + \xi_k + f_k(0)]} 
\ket{\varphi_k (0)} = \ket{\varphi_k (0)} . 
\end{eqnarray}
Here, $\ket{\varphi_k(t)}$ defines a cyclic lift 
of the evolution path in projective Hilbert space 
and can be viewed as a local section of the associated 
principal U(1) fiber bundle \cite{bohm91}. 

The eigenvectors represent cyclic states of the 
Schr\"odinger evolution $t \in [0,T] \mapsto U(t,0)$, 
in the sense that each $\ket{\psi_k}$ returns to 
its initial ray at $t=T$. The phase acquired during the 
cyclic evolution can be decomposed as 
$\xi_k = \gamma_k + \delta_k$, where
\begin{eqnarray}
\label{eq:geo_phase_ah-an}
\gamma_k & =  i 
\int_0^T\bra{\varphi_k(t)}\dot{\varphi}_k(t)\rangle dt,
\end{eqnarray} 
and 
\begin{eqnarray}
\label{eq:dyn_phase_ah-an}
\delta_k = - \frac{1}{\hbar} 
\int_{0}^{T}\bra{\varphi_k(t)}H(t)\ket{\varphi_k(t)}dt
\end{eqnarray}
are the geometric and dynamical contributions, 
respectively \cite{aharonov87}. 

We define a {\it geometric gate} as a unitary $U(T,0)$ 
for which all eigenstates accumulate the same dynamical 
phase, i.e., $\delta_k - \delta_1 = 0 \ \textrm{modulo} 
\ 2\pi$, $\forall k$ \cite{eivarsson23}. Universal 
geometric quantum computation requires unitary operators 
that are geometric with respect to different eigenbases 
of the computational subspace $\nu_0$. 

\subsection{Holonomic quantum gates}
\label{sec:def_HQC}
Consider a proper subspace $\nu(t)$ of Hilbert space 
$\mathscr{H}$ with $\dim\nu(t) = N<\dim\mathscr{H}$, 
undergoing a smooth cyclic evolution over the interval 
$t \in [0,T]$, i.e., $\nu(T)=\nu(0)$, generated by a 
time-dependent Hamiltonian $H(t)$. We associate this 
evolving subspace with a rank-$N$ projector 
$P(t) = U(t,0) P(0) U^{\dagger}(t,0)$. In contrast 
to the Abelian case discussed above, where the 
computational states remain confined to the initial 
subspace, the subspace $\nu (t)$ itself evolves along 
a path in the space of subspaces, such that 
$P(t) \neq P(t')$ for generic $t,t' \in [0,T]$. 
Consequently, an orthonormal basis of $\nu(0)$ will 
generally not consist of eigenstates of the final 
evolution operator $U(T,0)$. Instead, the basis 
states evolve 
into linear combinations of one another, while 
collectively spanning the initial subspace at the 
cyclic time $t=T$, i.e., $P(T)=P(0)$. The cyclic 
evolution induces a unitary transformation within 
$\nu(0)$. This projected unitary operator 
$\bar{U}(T,0) \equiv P(0) U(T,0) P(0)$ is 
the quantum gate operator acting on the computational
subspace $\nu(0)$. 

To describe the time evolution, consider a set of 
orthonormal vectors $\{\ket{\varphi_k(t)}\}_{k=1}^N$, 
each assumed to be periodic,
$\ket{\varphi_k(T)}=\ket{\varphi_k(0)}$, and spanning 
the evolving subspace $\nu(t)$. The collection of 
these vectors defines a smooth orthonormal moving 
frame, which can be regarded as a local section of
the Stiefel bundle over the Grassmannian of 
$N$-dimensional subspaces of $\mathscr{H}$. In this 
non-Abelian setting, the freedom in choosing
the local section is described by the unitary 
group $U(N)$, corresponding to changes of orthonormal 
basis within the evolving subspace.

We also consider a second orthonormal set
$\{\ket{\psi_k(t)}\}_{k=1}^N$, whose elements 
satisfy the Schr\"odinger equation 
$i\ket{\dot{\psi}_k(t)} = H(t) \ket{\psi_k(t)}$, 
with both sets spanning $\nu(t)$. Thus,
\begin{eqnarray}
P(t) = 
\sum_{j=1}^N \ket{\varphi_j(t)} \bra{\varphi_j(t)} = 
\sum_{j=1}^N \ket{\psi_j(t)}\bra{\psi_j(t)}.
\end{eqnarray}
The two orthonormal frames are therefore related 
by a time-dependent unitary transformation,
\begin{eqnarray}
\ket{\psi_k(t)} =
\sum_{j=1}^N \ket{\varphi_j(t)}\mathbb{U}_{jk}(t),
\end{eqnarray}
where the matrix elements of the $N\times N$ 
unitary matrix
$\mathbb{U}(t)$ are given by
\begin{eqnarray}
\mathbb{U}_{jk}(t) =
\langle\varphi_j(t)\ket{\psi_k(t)}.
\end{eqnarray}
This relation is the non-Abelian counterpart of 
the Abelian decomposition in 
Eq.~\eqref{eq:abeliansection}, with the phase 
factor $e^{if_k(t)}$ replaced by a unitary 
transformation in $U(N)$.

Since each $\ket{\psi_k(t)}$ satisfies the 
Schr\"odinger equation, the evolution of the 
transformation matrix is governed by the
non-Abelian geometric and dynamical contributions. 
In particular, at the end of the evolution, one  
obtains \cite{anandan88}
\begin{eqnarray}
\mathbb{U}(T) = \mathcal{T} \left( e^{i\int_0^T 
[\mathbb{A}(t) + \mathbb{K}(t)]dt}\right) , 
\end{eqnarray} 
where the two Hermitian $N\times N$ matrices 
$\mathbb{A}(t)$ and
$\mathbb{K}(t)$ are defined through their matrix 
elements as
\begin{eqnarray}
\label{eq:matrix_anandan_geo}
\mathbb{A}_{jk}(t) &=&
i\langle\varphi_j(t)\ket{\dot{\varphi}_k(t)},\\
\mathbb{K}_{jk}(t) &=&
-\frac{1}{\hbar}
\bra{\varphi_j(t)}H(t)\ket{\varphi_k(t)}.
\end{eqnarray}
Here, $\mathbb{A}(t)$ describes the non-Abelian 
geometric connection associated with the choice 
of moving frame, whereas $\mathbb{K}(t)$ describes 
the dynamical contribution projected onto the 
evolving subspace. Since the moving frame 
$\{\ket{\varphi_k(t)}\}_{k=1}^N$ 
is periodic, the matrix $\mathbb{U}(T)$ relates the 
Schr\"odinger-evolved basis states to the 
initial frame and determines the quantum 
gate acting on the computational subspace $\nu(0)$. 
Consequently, the gate operator can be expressed as
\begin{eqnarray}
\bar{U}(T,0) & = & \sum_{j,k=1}^{N} \mathbb{U}_{jk} (T)
\ket{\varphi_j(0)}\bra{\varphi_k(0)}.
\end{eqnarray}

If $\mathbb{K}_{jk}(t) \propto \delta_{jk}$, 
$\bar{U}(T,0) \equiv U(C)$ is said to be a 
{\it holonomic gate} associated with the loop 
$C:\: t\in[0,T] \mapsto \nu(t)$, $\nu(T)=\nu(0)$,  
in the space of $N$-dimensional subspaces of 
$\mathscr{H}$. Earlier work have demonstrated 
the existence of loops in physically realizable 
systems that are sufficiently flexible so as to 
guarantee universality \cite{sjoqvist12}. 

\section{Successive combinations of gates}
\label{sec:Successive combinations}

\subsection{Geometric quantum computation}
\label{sec:GQC_combinations}

The Zhu--Wang (ZW) scheme \cite{zhu03,zhu02} can 
be used as the basis for a geometric implementation 
of a phase-shift gate with respect to an orthonormal 
set $\{\ket{\psi_+},\ket{\psi_-}\}$, where
\begin{eqnarray}
\ket{\psi_+} & = & 
\cos\frac{\zeta}{2}\ket{0}
+e^{i\chi}\sin\frac{\zeta}{2}\ket{1},
\nonumber \\
\ket{\psi_-} & = &
-e^{-i\chi}\sin\frac{\zeta}{2}\ket{0}
+\cos\frac{\zeta}{2}\ket{1}.
\end{eqnarray}
The corresponding phase-shift operation is
\begin{eqnarray}
\ket{\psi_\pm}\longrightarrow
e^{\pm i\xi}\ket{\psi_\pm}.
\end{eqnarray}
When the dynamical phases vanish, the cyclic 
phases acquired by the two eigenstates 
$\ket{\psi_{\pm}}$ are purely geometric and 
are given by $\mp\frac{1}{2}\Omega$, respectively, 
where $\Omega$ is the oriented solid angle enclosed 
by the evolution path. Thus, $\xi = -\frac{1}{2}\Omega$ modulo $2\pi$.

Although individual gates can be implemented 
geometrically, their concatenation does not 
necessarily preserve this property. In particular, 
universal quantum computation requires noncommuting 
gates, which generally have different eigenvectors. 
Consequently, an eigenstate of the resulting composite 
gate need not coincide with those of its constituent 
gates, and nontrivial dynamical phases can accumulate 
during their successive implementation. 

To investigate this point, we consider the 
following two geometric gates:
\begin{eqnarray}
R_z(\Omega)=
\begin{pmatrix}
e^{-i\frac{1}{2}\Omega} & 0 \\
0 & e^{i\frac{1}{2}\Omega}
\end{pmatrix}
= e^{-i\frac{\Omega}{2}\sigma_z},
\end{eqnarray}
and
\begin{eqnarray}
R_x(\Omega) = \begin{pmatrix}
\cos\frac{\Omega}{2}
 & -i\sin\frac{\Omega}{2} \\
-i\sin\frac{\Omega}{2} & \cos\frac{\Omega}{2}
\end{pmatrix}
= e^{-i\frac{\Omega}{2}\sigma_x}.
\end{eqnarray}
These gates implement rotations through angles 
$\Omega$ and $\Omega$ about the $z$ and $x$ 
axes, respectively. Within the ZW scheme, they 
can be realized by choosing $\zeta=0$ with 
arbitrary $\chi$ for $R_z(\Omega)$, and 
$\zeta=\frac{\pi}{2}$ and $\chi=0$ for 
$R_x(\Omega)$.

Both geometric gates can be implemented using 
orange-slice paths on the Bloch sphere 
\cite{tian04,ota09a,thomas11}. For the $z$ 
rotation, a two-pulse Hamiltonian of the form 
\cite{remark1}
\begin{eqnarray}
H_z (t) = \left\{ \begin{array}{cl} 
\frac{1}{2} \hbar \omega \sigma_y, & 0 \leq t \leq 
\frac{\pi}{\omega} , \\[4pt]
\frac{1}{2} \hbar \omega (- \cos \phi \sigma_y + 
\sin \phi \sigma_x), & \frac{\pi}{\omega} 
\leq t \leq \frac{2\pi}{\omega} ,  
\end{array} \right. 
\nonumber \\  
\end{eqnarray}
generates geodesic paths along semicircles of great 
circles connecting the north and south poles of the 
Bloch sphere. The dihedral angle $\phi$ between the 
two great circles determines the enclosed oriented 
solid angle $\Omega = 2\phi$. Since the Hamiltonian 
is piecewise constant, the corresponding evolution 
operator can be explicitly evaluated as 
\begin{eqnarray}
U_z\left(\frac{2\pi}{\omega},0\right) & = & 
e^{-i\frac{\pi}{2} (-\cos\phi \, \sigma_y + 
\sin\phi \, \sigma_x)} e^{-i\frac{\pi}{2}\sigma_y} 
\nonumber \\
 & = & \cos\phi \, I-i\sin\phi \, \sigma_z = R_z(\Omega).
\end{eqnarray}
The $x$ rotation can be implemented by
\begin{eqnarray}
H_x (t) = \left\{
\begin{array}{cl}
\frac{1}{2} \hbar\omega\sigma_y,
& 0\leq t\leq \frac{\pi}{\omega}, \\[4pt]
\frac{1}{2} \hbar \omega \left( -\cos\phi 
\,\sigma_y - \sin\phi\,\sigma_z \right),
& \frac{\pi}{\omega} \leq t \leq \frac{2\pi}{\omega},
\end{array}
\right.
\nonumber \\ 
\end{eqnarray}
This Hamiltonian generates an orange-slice path 
connecting two antipodal points on the equator, 
corresponding to the mutually orthogonal states
$\ket{\pm} = \frac{1}{\sqrt{2}}(\ket{0}\pm\ket{1})$. 
The dihedral angle $\phi$ defines the enclosed 
oriented solid angle $\Omega=2\phi$. The 
resulting evolution operator is
\begin{eqnarray}
U_x\left(\frac{2\pi}{\omega},0\right) & = & 
e^{-i\frac{\pi}{2} (-\cos\phi \, \sigma_y -\sin\phi 
\, \sigma_z)} e^{-i\frac{\pi}{2}\sigma_y}
\nonumber \\
 & = & \cos\phi \, I-i\sin\phi \, \sigma_x =  
R_x(\Omega).
\end{eqnarray}

We now consider the composite gate
\begin{eqnarray}
U=R_z(\Omega_3)R_x(\Omega_2)R_z(\Omega_1),
\end{eqnarray}
where $\Omega_j$, $j=1,2,3$, play the role of 
Euler angles. This decomposition allows for the 
exact implementation of an arbitrary single-qubit 
unitary. Together with an entangling two-qubit 
gate, it therefore provides {\it exact} universality 
for quantum computation \cite{brylinski02}. 
Although each constituent rotation is implemented 
geometrically, the complete sequence may acquire 
a non-zero dynamical phase. We next evaluate this 
phase directly in terms of the Bloch vector of 
an eigenstate of $U$.

For a constant Hamiltonian $H = \frac{1}{2} 
\hbar \omega \, \mathbf{n} \cdot \boldsymbol{\sigma}$ 
and time $t$, the Bloch vector rotates about 
$\mathbf{n}$ through an angle $\omega t$. Since each 
pulse has duration $\frac{\pi}{\omega}$, the 
corresponding Bloch-sphere rotation angle is $\pi$, 
and the projection of the Bloch vector onto the 
rotation axis remains constant throughout each 
pulse. Thus, the dynamical-phase contribution of 
a pulse is $-\frac{\pi}{2} \mathbf{n} \cdot 
\mathbf{r}_{\mathrm{in}}$, where 
$\mathbf{r}_{\mathrm{in}}$ denotes the Bloch 
vector at the beginning of that pulse.

For both orange-slice implementations, the first 
pulse rotates the initial Bloch vector $\mathbf{r}_0$ 
according to
\begin{eqnarray}
\mathbf{r}_a = R_y(\pi)\mathbf{r}_0
= (-r_{0x},r_{0y},-r_{0z}).
\end{eqnarray}
For the $z$ rotation, the two rotation axes are 
$\mathbf{n}_1=(0,1,0)$ and $\mathbf{n}_2 = 
(\sin\phi,-\cos\phi,0)$. Adding the dynamical-phase 
contributions from the two pulses therefore gives
\begin{eqnarray}
\delta_z=\frac{\pi}{2}\left[
r_{0x}\sin\phi+r_{0y}(\cos\phi-1)
\right].
\end{eqnarray}
Using $\Omega = 2\phi$, this becomes
\begin{eqnarray}
\delta_z=\frac{\pi}{2}\left[
r_{0x}\sin\frac{\Omega}{2}
+r_{0y}\left(\cos\frac{\Omega}{2}-1\right)
\right].
\end{eqnarray}
Similarly, for the $x$ rotation, the two 
rotation axes are $\mathbf{n}_1 = (0,1,0)$ and 
$\mathbf{n}_2 = (0,-\cos\phi_2,-\sin\phi_2)$, 
giving
\begin{eqnarray}
\delta_x = \frac{\pi}{2}\left[
r_{0y} (\cos\phi_2-1) -r_{0z} \sin\phi_2
\right].
\end{eqnarray}
Setting $\Omega_2 = 2\phi_2$ gives
\begin{eqnarray}
\delta_x = \frac{\pi}{2}\left[
r_{0y}\left(\cos\frac{\Omega_2}{2}-1\right)
-r_{0z}\sin\frac{\Omega_2}{2}
\right].
\end{eqnarray}

For the composite gate $U$, the dynamical-phase 
contributions of the three successive rotations 
can be added directly. Thus, for an initial Bloch 
vector $\mathbf{r}_0$, the total dynamical phase is
\begin{eqnarray}
\delta_{\mathrm{tot}} & = & \frac{\pi}{2}\bigg[
r_{0x}\sin\frac{\Omega_1}{2}
+r_{0y}\left(\cos\frac{\Omega_1}{2}-1\right)
\nonumber\\
& & +r_{1y}\left(\cos\frac{\Omega_2}{2}-1\right)
-r_{1z}\sin\frac{\Omega_2}{2}
\nonumber\\
& & +r_{2x}\sin\frac{\Omega_3}{2}
+r_{2y}\left(\cos\frac{\Omega_3}{2}-1\right)
\bigg],
\label{eq:total_dynamical_phase_expanded}
\end{eqnarray}
where the Bloch vectors at the boundaries between 
successive gates are
\begin{eqnarray}
\mathbf{r}_1=R_z(\Omega_1)\mathbf{r}_0, \qquad
\mathbf{r}_2=R_x(\Omega_2)\mathbf{r}_1.
\end{eqnarray}
Under the first rotation $R_z(\Omega_1)$, the Bloch 
vector transforms as
\begin{eqnarray}
r_{1x} & = & r_{0x}\cos\Omega_1-r_{0y}\sin\Omega_1, 
\nonumber \\
r_{1y} & = & r_{0x}\sin\Omega_1+r_{0y}\cos\Omega_1,
\nonumber \\
r_{1z} & = & r_{0z}. 
\label{eq:bloch_vector_after_Rz1}
\end{eqnarray}
The subsequent rotation $R_x(\Omega_2)$ gives
\begin{eqnarray}
r_{2x} & = & r_{1x},
\nonumber \\
r_{2y} & = & r_{1y}\cos\Omega_2-r_{1z}\sin\Omega_2,
\nonumber \\
r_{2z} & = & r_{1y}\sin\Omega_2+r_{1z}\cos\Omega_2.
\label{eq:bloch_vector_after_Rx}
\end{eqnarray}
Since the initial state is an eigenstate of $U$, its 
Bloch vector returns to itself after the three 
rotations. The cyclicity condition 
$\mathbf{r}_3=\mathbf{r}_0$ therefore gives
\begin{eqnarray}
r_{0x} & = & r_{2x}\cos\Omega_3-r_{2y}\sin\Omega_3,
\nonumber \\
r_{0y} & = & r_{2x}\sin\Omega_3+r_{2y}\cos\Omega_3,
\nonumber \\
r_{0z} & = & r_{2z}.
\label{eq:cyclicity_condition}
\end{eqnarray}
Using Eqs.~\eqref{eq:bloch_vector_after_Rz1}--\eqref{eq:cyclicity_condition}, the total 
dynamical phase in
Eq.~\eqref{eq:total_dynamical_phase_expanded} 
can be simplified to
\begin{eqnarray}
\delta_{\mathrm{tot}} = \frac{\pi}{4} \left[ 
\tan \frac{\Omega_1}{4} - \tan\frac{\Omega_3}{4} \right]
(r_{0x}-r_{2x}).
\label{eq:total_dynamical_phase_general}
\end{eqnarray}
By combining Eqs.~\eqref{eq:bloch_vector_after_Rz1} 
and \eqref{eq:bloch_vector_after_Rx}, we find 
\begin{eqnarray}
r_{2x}=r_{1x}
=r_{0x}\cos\Omega_1-r_{0y}\sin\Omega_1.
\end{eqnarray}
Substituting this relation into 
Eq.~\eqref{eq:total_dynamical_phase_general}, 
we finally obtain
\begin{eqnarray}
\delta_{\mathrm{tot}} & = & 
\frac{\pi}{4} \left[ \tan\frac{\Omega_1}{4} -
\tan\frac{\Omega_3}{4} \right]
\nonumber\\
 & &\times \left[ r_{0x}(1-\cos\Omega_1) + r_{0y}
\sin\Omega_1
\right].
\label{eq:total_dynamical_phase_explicit}
\end{eqnarray}
For the other eigenstate of $U$, whose initial Bloch 
vector is $-\mathbf{r}_0$, the dynamical phase changes 
sign. Consequently,
\begin{eqnarray}
\delta_{\mathrm{tot},-} = -\delta_{\mathrm{tot},+}.
\end{eqnarray}

An interesting special case occurs when either 
$\Omega_1=0$ or $\Omega_3=0$, for which the total 
dynamical phase vanishes independently of the value 
of the remaining rotation angles. While the former 
follows directly from
Eq.~\eqref{eq:total_dynamical_phase_explicit}, the 
latter follows from the cyclicity condition in 
Eq.~\eqref{eq:cyclicity_condition}, which 
for $\Omega_3=0$ yields $r_{2x}=r_{0x}$. Consequently, 
the dynamical-phase contributions from the remaining two 
rotations cancel exactly, irrespective of $\Omega_2$. 
This cancellation does not arise from 
parallel transport during the individual rotations, 
but rather from the cancellation of their accumulated 
dynamical phases over the complete sequence. This is 
an example of how the accumulated dynamical phases can 
be made to vanish through an appropriate choice of 
the gate parameters, as discussed in the introduction 
\cite{zhu03,ota09b}. The cancellation further 
illustrates that noncommutativity of the constituent 
gates, although necessary for the possibility of 
generating a nonzero dynamical phase in this setting, 
is not sufficient. Nevertheless, implementing an 
arbitrary single-qubit unitary through the full 
Euler-angle decomposition generally requires all 
three rotations, and thus introduces nontrivial 
dynamical phases for generic choices of Euler angles.

The above analysis shows that the total dynamical 
phase is generally non-zero. Thus, although each 
individual gate is purely geometric, their 
concatenation does not generally preserve this 
property. This illustrates explicitly how the 
geometric character of individual gates does not 
necessarily extend to a concatenated operation, 
when the constituent gates do not share a common 
eigenbasis.

\subsection{Holonomic quantum computation}
\label{sec:HQC_combinations}

The single-loop multiple-pulse scheme introduced 
in Ref.~\cite{herterich16} is an efficient tool 
for implementing holonomic gates in a three-level 
$\Lambda$ system. In this scheme, two resonant 
laser fields couple an auxiliary excited state 
$\ket{e}$ to the bright state
\begin{eqnarray}
\ket{b}=\omega_0^*\ket{0}+\omega_1^*\ket{1},
\end{eqnarray}
while leaving the orthogonal dark state 
$\ket{d}=-\omega_1\ket{0}+\omega_0\ket{1}$ 
decoupled, where $|\omega_0|^2+|\omega_1|^2=1$. 
The states $\ket{0}$  and $\ket{1}$ span the 
computational subspace $\nu (0)$. The states 
$\ket{0},\ket{1}$, and $\ket{e}$ are mutually 
orthogonal. The evolution consists of two consecutive 
resonant pulse segments, governed by
\begin{eqnarray}
H^{(i)}(t) & = & \hbar F^{(i)}(t)
(\ket{e}\bra{b}+\ket{b}\bra{e}),
\nonumber\\
H^{(ii)}(t) & = & \hbar F^{(ii)}(t)
(e^{-i\eta}\ket{b}\bra{e}+e^{i\eta}\ket{e}\bra{b}),
\end{eqnarray}
respectively, where $\eta$ is the relative phase shift 
between the segments. These Hamiltonians parallel 
transport the evolving computation subspace $\nu(t) = 
\textrm{Span} \{ \ket{d},U(t,0)\ket{b} \}$. 
Choosing both pulse areas equal to $\frac{\pi}{2}$ 
yields the holonomic gate 
\begin{eqnarray}
U_{\mathrm{HQC}}=\ket{d}\bra{d}-e^{-i\eta}\ket{b}\bra{b}
\end{eqnarray}
acting on the computational subspace 
$\nu (0)$ since $\textrm{Span} \{ \ket{d},\ket{b} \} = 
\textrm{Span} \{ \ket{0},\ket{1} \}$. Equivalently, 
this corresponds to a Bloch-sphere rotation
\begin{eqnarray}
U_{\mathrm{HQC}} & = & e^{i\frac{1}{2} (\pi-\eta)} 
\exp\left[-i\frac{\pi-\eta}{2} \mathbf{n} 
\cdot\boldsymbol{\sigma}\right]
\nonumber \\ 
 & = & e^{i\frac{1}{2} (\pi-\eta)} 
R_{\mathbf{n}} (\pi-\eta) ,
\end{eqnarray}
with
\begin{eqnarray}
\mathbf{n} = 
\begin{pmatrix}
-2\mathrm{Re}(\omega_0\omega_1^*)\\
-2\mathrm{Im}(\omega_0\omega_1^*)\\
|\omega_1|^2-|\omega_0|^2
\end{pmatrix}.
\end{eqnarray}
Thus, the laser amplitudes determine the rotation 
axis $\mathbf{n}$, while their relative phase shift 
determines the rotation angle $\pi-\eta$, allowing 
arbitrary single-qubit gates to be realized within 
a single loop.

To facilitate comparison with the Abelian case, 
we consider the corresponding Euler-angle 
decomposition: 
\begin{eqnarray}
U=R_z(\pi-\eta_3)R_x(\pi-\eta_2)R_z(\pi-\eta_1),
\label{eq:HQC_Euler_gate}
\end{eqnarray}
where
\begin{eqnarray}
R_z(\pi-\eta_j)& = & 
e^{-i\frac{1}{2}(\pi-\eta_j)\sigma_z},
\nonumber\\
R_x(\pi-\eta_2)& = & 
e^{-i\frac{1}{2}(\pi-\eta_2)\sigma_x}.
\end{eqnarray}
Since $U\in\mathrm{SU}(2)$, it can be expressed in 
the axis-angle representation
\begin{eqnarray}
U=\cos\frac{\theta_U}{2}I
-i\sin\frac{\theta_U}{2} 
\mathbf{n}_U\cdot\boldsymbol{\sigma},
\label{eq:HQC_Euler_axis_angle}
\end{eqnarray}
where $\theta_U$ and $\mathbf{n}_U$ are the rotation 
angle and unit rotation axis, respectively. They are 
determined by
\begin{eqnarray}
\cos\frac{\theta_U}{2} & = & 
\frac{1}{2}\mathrm{Tr}(U),
\nonumber \\
\mathbf{n}_U & = & 
\frac{i}{2\sin\frac{\theta_U}{2}}
\mathrm{Tr}(\boldsymbol{\sigma}U),
\label{eq:single_parameters}
\end{eqnarray}
for $\sin \frac{\theta_U}{2}\neq0$.

One may distinguish between implementing the Euler-angle 
operation through three successive single-loop holonomic 
gates and implementing the complete operation using a 
single-loop multiple-pulse construction. In the former 
case, each individual rotation contributes a global 
phase $\frac{1}{2}(\pi-\eta_j)$, $j=1,2,3$. 
Consequently, the concatenated holonomic operation is
\begin{eqnarray}
U_{\mathrm{HQC}}^{(3)} =
e^{i\frac{1}{2}(3\pi-\eta_1-\eta_2-\eta_3)}U,
\label{eq:UHQC3}
\end{eqnarray}
whereas a single-loop realization of the complete 
operation gives
\begin{eqnarray}
U_{\mathrm{HQC}}^{(1)}=e^{i\frac{1}{2} \theta_U}U 
\end{eqnarray}
with $\theta_U$ given in 
Eq.~\eqref{eq:single_parameters}. The two 
implementations therefore differ by the global phase
\begin{eqnarray}
\Delta\xi=
\frac{1}{2}(3\pi-\eta_1-\eta_2-\eta_3-\theta_U),
\end{eqnarray}
which is generally non-zero. Although this 
difference is unobservable, it demonstrates that 
the concatenation of holonomic gates does not 
necessarily yield the same unitary operator as 
a direct implementation of their product.

The phase difference $\Delta\xi$ provides an 
interesting point of comparison with the dynamical 
phases arising from successive geometric gates in 
the ZW scheme. In both cases, the phase associated 
with a concatenated implementation differs from that 
obtained by considering the complete operation as 
a single gate. However, the physical consequences 
of these phase differences are fundamentally distinct. 
In the ZW scheme, the accumulated dynamical phases 
generally have opposite signs for the two eigenstates 
of the composite gate, thereby producing a relative 
phase that modifies the resulting quantum operation. 
Consequently, the concatenated evolution is generally 
no longer purely geometric. In contrast, the phase
difference $\Delta\xi$ in the holonomic scheme is a 
global phase multiplying the entire computational-space 
operation. It therefore has no effect on the action 
of the gate on a computational state. In particular, 
it does not 
signify the emergence of a dynamical phase or a 
breakdown of the parallel-transport condition. 
Rather, it reflects the fact that concatenating
several holonomic loops and implementing their 
combined operation through a single loop can yield 
different overall geometric phases, even when the 
resulting SU(2) operations are identical.

This distinction can also be understood by considering 
the difference between SU(2) and U(2) operations. 
Although the Euler-angle decomposition describes 
an operation $U\in\mathrm{SU}(2)$, the individual 
holonomic gates generally belong to U(2), as 
reflected by their non-unit determinants. 
Consequently, their global phases accumulate upon
concatenation, whereas a direct single-loop 
implementation assigns a global phase determined 
by the axis-angle representation of the resulting 
SU(2) operation. The difference $\Delta\xi$ thus 
reflects the distinct global geometric phases 
associated with two different implementations
of the same computational operation, rather than a
modification of the operation itself. This is in 
marked contrast to the ZW scheme, where the 
dynamical phases generated by concatenation 
generally modify the relative phases of the 
computational eigenstates and hence the resulting 
gate. Thus, while concatenation in the ZW scheme
can compromise the purely geometric character of 
the evolution, the holonomic construction 
considered here preserves this character and 
changes only an overall phase.

Although the three constituent gates generally have 
different bright--dark bases, their successive 
application preserves the purely geometric character of 
the evolution. To establish this property, we examine 
the action of the corresponding Hamiltonians on an 
arbitrary state in the computational subspace 
$\nu (0) = \textrm{Span} \{ \ket{0},\ket{1} \}$. 
We will subsequently show that each elementary loop 
can also be understood geometrically as an Abelian 
orange-slice phase gate in the bright--excited 
subspace.

For the three successive primitve gates in 
Eq.~\eqref{eq:HQC_Euler_gate}, we choose the 
bright and dark states as
\begin{eqnarray}
\ket{b_1} & = & \ket{b_3}=\ket{1},
\nonumber\\
\ket{d_1} & = & \ket{d_3}=-\ket{0},
\end{eqnarray}
for the first and third rotations, and
\begin{eqnarray}
\ket{b_2} & = & \ket{-}
=\frac{1}{\sqrt{2}}(\ket{0}-\ket{1}),
\nonumber\\
\ket{d_2} & = & -\ket{+}
=-\frac{1}{\sqrt{2}}(\ket{0}+\ket{1}),
\end{eqnarray}
for the second rotation. The corresponding 
Hamiltonians can be written in a common form:
\begin{eqnarray}
H_j^{(i)}(t)=\hbar F_j^{(i)}(t)
\left(\ket{e}\bra{b_j}+\ket{b_j}\bra{e}\right),
\end{eqnarray}
and
\begin{eqnarray}
H_j^{(ii)}(t)=\hbar F_j^{(ii)}(t)
\left(
e^{-i\eta_j}\ket{b_j}\bra{e}
+e^{i\eta_j}\ket{e}\bra{b_j}
\right),
\end{eqnarray}
where $j=1,2,3$. We denote the two pulse areas by
\begin{eqnarray}
a_j^{(i)} & = & \int_0^\tau F_j^{(i)}(t)dt,
\nonumber\\
a_j^{(ii)} & = & \int_{\tau}^T F_j^{(ii)}(t)dt.
\end{eqnarray}
For the ideal holonomic construction considered below,
both pulse areas are chosen equal to $\frac{\pi}{2}$.
To investigate the role of this choice in ensuring
purely geometric evolution, we will also consider
more general pulse areas.

The essential feature of these Hamiltonians is that 
they couple only the bright state to the  
excited state, while the dark state remains 
decoupled. For the ideal holonomic construction, the 
relative phase between the bright and excited state 
amplitudes remains such that the expectation value 
of the coupling Hamiltonian vanishes instantaneously 
throughout both pulse segments. This cancellation 
holds for an arbitrary superposition of bright and 
dark states, rather than only for a particular 
eigenstate of the gate.

To establish this property, consider an arbitrary 
normalized initial state in the computational 
subspace at the beginning of the $j$th loop,
\begin{eqnarray}
\ket{\Psi_{j-1}} =
\alpha_j\ket{d_j}+\beta_j\ket{b_j},
\label{eq:HQC_bright_dark_decomposition}
\end{eqnarray}
where $|\alpha_j|^2+|\beta_j|^2=1$. During the 
first pulse, the state evolves as
\begin{eqnarray}
\ket{\Psi_j(t)}
&=&\alpha_j\ket{d_j}
\nonumber\\
&&+\beta_j\left[
\cos f_j^{(i)}(t)\ket{b_j}
-i\sin f_j^{(i)}(t)\ket{e}
\right],
\end{eqnarray}
with
\begin{eqnarray}
f_j^{(i)}(t)=
\int_0^t F_j^{(i)}(t')dt'.
\end{eqnarray}
The relative phase factor $-i$ between the 
bright and excited state amplitudes consequently gives
\begin{eqnarray}
 & & \bra{\Psi_j(t)}H_j^{(i)}(t)\ket{\Psi_j(t)}
\nonumber \\ 
 & = & 2\hbar F_j^{(i)}(t)
\mathrm{Re}\left[
-i|\beta_j|^2
\sin f_j^{(i)}(t)\cos f_j^{(i)}(t)
\right] 
\nonumber\\
 & = & 0.
\label{eq:HQC_first_pulse_expectation}
\end{eqnarray}
For the pulse area $a_j^{(i)}=\frac{\pi}{2}$, 
the state at the end of the first pulse is 
\begin{eqnarray}
\ket{\Psi_j(\tau)}
=\alpha_j\ket{d_j}-i\beta_j\ket{e}.
\end{eqnarray}
During the second pulse, it evolves according to
\begin{eqnarray}
\ket{\Psi_j(t)}
&=&\alpha_j\ket{d_j}
-\beta_j e^{-i\eta_j}
\sin f_j^{(ii)}(t)\ket{b_j}
\nonumber\\
&&-i\beta_j\cos f_j^{(ii)}(t)\ket{e},
\label{eq:HQC_second_segment_state}
\end{eqnarray}
where
\begin{eqnarray}
f_j^{(ii)}(t)=
\int_{\tau'}^t F_j^{(ii)}(t')dt'.
\end{eqnarray}
The bright and excited state amplitudes again 
have a relative phase factor $-i$, with the 
additional laser phase $\eta_j$, which implies 
\begin{eqnarray}
 & & \bra{\Psi_j(t)}H_j^{(ii)}(t)\ket{\Psi_j(t)}
\nonumber \\ 
 & = & 2\hbar F_j^{(ii)}(t)
\mathrm{Re}\left[
i|\beta_j|^2
\sin f_j^{(ii)}(t)\cos f_j^{(ii)}(t)
\right]
\nonumber\\
 & = & 0.
\label{eq:HQC_second_pulse_expectation}
\end{eqnarray}
Thus, the instantaneous dynamical phase vanishes 
throughout both pulse segments for every initial 
state in the computational subspace. This establishes 
the parallel-transport condition independently of the 
initial bright-state population and, consequently, 
independently of the particular bright--dark basis 
used for the loop.

The role of the ideal pulse areas can also be 
seen by considering a general choice of $a_j^{(i)}$ 
and $a_j^{(ii)}$. In this case, the
expectation value of the second-pulse Hamiltonian is
\begin{eqnarray}
 & & \bra{\Psi_j(t)}H_j^{(ii)}(t)\ket{\Psi_j(t)}
\nonumber \\ 
 & = & -\hbar F_j^{(ii)}(t)|\beta_j|^2
\sin(2a_j^{(i)})\sin\eta_j,
\end{eqnarray}
giving a total dynamical phase
\begin{eqnarray}
\delta_j^{(ii)}
&=&-\frac{1}{\hbar}\int_{\tau'}^T
\bra{\Psi_j(t)}H_j^{(ii)}(t)\ket{\Psi_j(t)}dt
\nonumber\\
&=&|\beta_j|^2a_j^{(ii)}\sin(2a_j^{(i)})\sin\eta_j.
\label{eq:HQC_general_dynamic_phase}
\end{eqnarray}
Moreover, the final excited-state amplitude is generally
nonzero, as follows from
\begin{eqnarray}
 & & \ket{\Psi_j(T)} = \alpha_j\ket{d_j}
\nonumber \\
 & + & \beta_j\left[
(\cos a_j^{(i)}\cos a_j^{(ii)}
-e^{-i\eta_j}\sin a_j^{(i)}\sin a_j^{(ii)})\ket{b_j} \right. 
\nonumber\\
& - &\left. i(\sin a_j^{(i)}\cos a_j^{(ii)}
+e^{i\eta_j}\cos a_j^{(i)}\sin a_j^{(ii)})\ket{e}
\right].
\end{eqnarray}
Consequently, for general pulse areas, the evolution 
need not be cyclic with respect to the computational 
subspace, and a nonzero dynamical phase can arise. 
Note that, although a nonzero dynamical phase 
necessarily implies non-cyclic evolution when 
$|\beta_j|^2\sin\eta_j \neq 0$, the converse does 
not hold: non-cyclic evolution can occur without 
an accompanying dynamical phase if the first pulse
satisfies the ideal condition
$a_j^{(i)}=\frac{\pi}{2}$. Thus, while nonzero 
dynamical phase and non-cyclic evolution are related, 
they are distinct features of the evolution in this 
system.

For the ideal pulses where $a_j^{(i)}=a_j^{(ii)}=\frac{\pi}{2}$, 
the final state reduces to
\begin{eqnarray}
\ket{\Psi_j(T)}
&=&\alpha_j\ket{d_j}
-e^{-i\eta_j}\beta_j\ket{b_j}
\nonumber\\
&=&U_{\mathrm{HQC},j}\ket{\Psi_{j-1}}
\in\nu(0).
\label{eq:final_state_hol}
\end{eqnarray}
Hence, each ideal holonomic loop returns the 
computational subspace to itself, ensuring that 
its output is a valid initial state for the 
subsequent loop. Since the dynamical-phase
cancellation holds for every state in the 
computational subspace, it remains valid 
irrespective of the bright--dark basis chosen 
in successive loops. In particular, this applies
to the three gates in the Euler decomposition, 
for which $\mathrm{Span}\{\ket{d_j},\ket{b_j}\}
=\mathrm{Span}\{\ket{0},\ket{1}\}$,
$\forall j=1,2,3$.

A holonomic loop admits an alternative geometric 
interpretation in terms of an Abelian geometric 
phase gate as follows. Each elementary loop can be 
viewed as an orange-slice evolution on the Bloch 
sphere of the effective two-level subsystem spanned 
by the bright and excited states, 
$\{\ket{b_j},\ket{e}\}$. This provides a direct 
connection between the holonomic construction and 
the geometric phase gates discussed in the previous 
subsection.

To make this connection explicit, we introduce the 
Pauli operators associated with the effective 
two-level subsystem:
\begin{eqnarray}
\tau_{j,x} & = & \ket{b_j}\bra{e} + 
\ket{e}\bra{b_j}, 
\nonumber \\
\tau_{j,y} & = & -i\ket{b_j}\bra{e} + 
i\ket{e}\bra{b_j}, 
\nonumber \\ 
\tau_{j,z} & = & \ket{b_j}\bra{b_j} - 
\ket{e}\bra{e}. 
\end{eqnarray}
In this representation, the two-pulse Hamiltonians 
become
\begin{eqnarray}
H_j^{(i)}(t)
&=& \hbar F_j^{(i)}(t)\tau_{j,x},
\nonumber \\
H_j^{(ii)}(t)
&=& \hbar F_j^{(ii)}(t)
\left[
\cos\eta_j\,\tau_{j,x}
+\sin\eta_j\,\tau_{j,y}
\right].
\end{eqnarray}
Thus, the two pulse segments correspond to rotations 
about two different axes in the equatorial plane of 
the effective Bloch sphere. Since both pulse areas 
are $\frac{\pi}{2}$ in the ideal case, each segment 
generates a Bloch-sphere rotation through an angle 
$\pi$. Starting from the bright state $\ket{b_j}$, 
the first pulse takes the state from the north pole 
to the south pole, corresponding to $\ket{e}$, while 
the second pulse returns it to the north pole. The 
two trajectories are geodesic semicircles along 
different meridians and together form an 
orange-slice loop.

The relative laser phase $\eta_j$ determines the 
dihedral angle between the two meridians. The 
enclosed oriented solid angle is
\begin{eqnarray}
\Omega_j^{(\mathrm{eff})} = -2(\pi-\eta_j) .
\end{eqnarray}
According to the geometric-phase relation for a spin-$1/2$ system, the phase acquired by the bright state is therefore
\begin{eqnarray}
\xi_{b_j} = -\frac{1}{2}\Omega_j^{(\mathrm{eff})}
=\pi-\eta_j \quad \mathrm{modulo}\ 2\pi.
\end{eqnarray}
Indeed, this agrees with the phase obtained directly 
from the evolution under the two pulse segments:
\begin{eqnarray}
\ket{b_j} \longrightarrow -e^{-i\eta_j}\ket{b_j} =
e^{i(\pi-\eta_j)}\ket{b_j}, 
\end{eqnarray}
as seen in Eq.~\eqref{eq:final_state_hol}. 
Since the dynamical phase vanishes throughout the 
evolution for the chosen pulse areas $a_j^{(i)} = 
a_j^{(ii)} = \frac{\pi}{2}$, the acquired phase is 
entirely geometric. By contrast, the dark state 
$\ket{d_j}$ is decoupled from the Hamiltonian and 
remains unchanged. The resulting operation can 
consequently be expressed as
\begin{eqnarray}
U_{\mathrm{HQC},j} = \ket{d_j}\bra{d_j} +
e^{-i\frac{1}{2} \Omega_j^{(\mathrm{eff})}} 
\ket{b_j}\bra{b_j}.
\label{eq:HQC_abelian_GP}
\end{eqnarray}
Thus, each elementary holonomic gate is an Abelian 
geometric phase operation in the bright--dark basis: 
the bright and dark states are its eigenstates, 
and the relative phase between them is determined 
by the solid angle enclosed by the orange-slice 
path in the bright--excited subsystem. Equivalently, 
after extracting a global phase, the two computational 
eigenstates acquire opposite phase shifts of magnitude 
$\frac{1}{4} \Omega_j^{(\mathrm{eff})}$.

The non-adiabatic geometric-phase sum rule of
Ref.~\cite{fredriksson26b,orion26} provides an additional
perspective on this construction. In applying the sum
rule, it is essential to consider the complete physical
Hilbert space rather than only the computational
subspace. For the present three-level system, the
complete orthonormal basis is
$\{\ket{d_j},\ket{b_j},\ket{e}\}$.
The evolution over one elementary loop returns all
three basis states to their initial rays. Indeed,
the full evolution operator can be written as
\begin{eqnarray}
U_j^{(3)}
&=&
\ket{d_j}\bra{d_j}
-e^{-i\eta_j}\ket{b_j}\bra{b_j}
\nonumber\\
&&
-e^{i\eta_j}\ket{e}\bra{e}.
\label{eq:HQC_full_evolution}
\end{eqnarray}
Since the dynamical phase vanishes throughout the
evolution, the geometric phases associated with these
three cyclic states are
\begin{eqnarray}
\gamma_{d_j}&=&0,\nonumber\\
\gamma_{b_j}&=&\pi-\eta_j,\nonumber\\
\gamma_e&=&\pi+\eta_j,
\end{eqnarray}
where the phases are understood modulo $2\pi$.
Consequently,
\begin{eqnarray}
\gamma_{d_j}+\gamma_{b_j}+\gamma_e=0
\quad \mathrm{modulo}\ 2\pi,
\label{eq:HQC_geometric_phase_sum_rule}
\end{eqnarray}
in agreement with the non-adiabatic geometric-phase
sum rule.

Notice that the determinant of the holonomic gate
restricted to the computational subspace is
\begin{eqnarray}
\det U_{\mathrm{HQC},j}
=e^{i(\pi-\eta_j)},
\end{eqnarray}
which is generally different from unity. This does
not contradict the sum rule, since the computational
subspace is a proper subspace of the physical
three-dimensional Hilbert space. When the auxiliary
excited state is included, the full evolution operator
has unit determinant, with its geometric phase
compensating the total phase acquired by the
computational states. 

This interpretation also clarifies how successive 
holonomic gates can generate noncommuting single-qubit 
operations. Although each individual loop is an 
Abelian geometric phase operation, its bright and 
dark eigenstates depend on the laser amplitudes and 
can therefore change between successive loops. For 
the Euler decomposition considered above, the first 
and third gates have $\ket{b_1}=\ket{b_3}=\ket{1}$, 
whereas the second gate has $\ket{b_2}=\ket{-}$. 
Consequently, the corresponding geometric phase 
gates generally have different eigenbases and do 
not commute. Their concatenation can therefore generate 
arbitrary single-qubit operations, even though each 
elementary operation is associated with an Abelian 
geometric phase. Thus, noncommutativity at the 
computational level does not require non-Abelian 
geometric evolution within each elementary loop. 
Instead, it arises from the concatenation of Abelian 
geometric phase gates associated with different 
bright--dark bases, with the auxiliary excited 
state providing the effective two-level subsystem 
in which each geometric phase is accumulated.

The preceding analysis highlights an important 
distinction between the ZW and holonomic schemes 
regarding the origin of dynamical phases in successive 
gate operations. In the ZW scheme, nonzero dynamical 
phases can arise even when all constituent gates are 
implemented ideally, owing to the different eigenbases 
associated with successive noncommuting gates. Their 
absence in the ideal case relies on the semicircular 
geodesic arcs of the orange-slice construction, and 
deviations from this geometry can give rise to 
dynamical phases even at the level of individual 
gates. In contrast, for the holonomic scheme 
considered here, the parallel-transport condition 
is satisfied for arbitrary initial states in the 
computational subspace, independently of the 
bright--dark basis, provided that the ideal 
pulse-area conditions are fulfilled. Thus, unlike 
in the ZW scheme, any non-zero dynamical phases 
in successive holonomic gates originate from 
deviations from the ideal pulse-area conditions 
rather than from the noncommutativity of the 
gates themselves.

\section{Conclusions}
\label{sec:Conclusions}

The possibility of realizing identical unitary 
transformations using Abelian geometric and 
non-Abelian holonomic schemes makes the distinction 
between their respective geometric properties 
particularly relevant in the context of concatenated 
quantum operations. In this work, we have investigated 
the geometric properties of concatenated quantum 
gates, focusing on the differences between geometric 
and holonomic quantum computation. We have shown that 
the geometric character of individual gates based on 
Abelian geometric phases does not, in general, extend 
to their successive application. This occurs because 
the cyclic states of a composite operation need not 
coincide with the cyclic states of its constituent 
gates, so that nontrivial dynamical phases can arise 
even when each individual gate is implemented 
geometrically. Our analysis of a sequence of 
Euler-angle rotations provides an explicit 
illustration of this effect.

In contrast, we have demonstrated that non-adiabatic 
holonomic quantum gates preserve their purely 
geometric character under concatenation, provided 
that each constituent gate satisfies the 
parallel-transport condition and returns the 
computational subspace to itself. In particular, 
for the single-loop multiple-pulse scheme introduced 
in Ref.~\cite{herterich16}, the ideal pulse-area 
conditions ensure that the instantaneous expectation 
value of the Hamiltonian vanishes for every state 
in the computational subspace, irrespective of the 
bright--dark basis associated with the particular loop. 
Consequently, arbitrary single-qubit operations can 
be constructed through successive applications of 
such gates without accumulating dynamical phases. 
This state-independent cancellation distinguishes 
the holonomic construction considered here from the 
Abelian ZW scheme, where dynamical-phase cancellation 
generally depends on the state and is typically lost 
when noncommuting gates are concatenated.

This distinction is important when considering the 
potential robustness of quantum circuits constructed 
from geometric and holonomic gates. Since arguments 
for geometric protection can only be extended 
consistently to complete quantum circuits when the 
relevant geometric conditions are preserved under 
concatenation, establishing these conditions is 
essential. In particular, our results show that 
the robustness arguments associated with individual 
geometric gates cannot automatically be extended to 
circuits composed of such gates. For holonomic 
circuits, the state-independent parallel-transport 
condition provides a sufficient condition for 
maintaining the geometric character of the complete 
operation under successive 
ideal gates. Nevertheless, robustness against 
experimental imperfections must be assessed 
separately, since deviations that alter the 
evolution paths, pulse areas, or cyclicity 
conditions can compromise the ideal holonomic
construction.

Our findings can also be viewed in the broader 
context of the non-adiabatic geometric-phase sum 
rule \cite{fredriksson26b}, according to 
which the geometric phases accumulated by a complete 
orthonormal basis of cyclic states sum to zero 
modulo $2\pi$ \cite{orion26}. We have provided 
explicit illustrations of this constraint in both 
geometric and holonomic quantum computation. For 
the concatenated ZW gates, the geometric phases 
of the two cyclic eigenstates of the composite 
operation compensate each other, even when their 
individual dynamical phases are nonzero. For the 
holonomic gates, the corresponding compensation 
involves the auxiliary excited state, emphasizing 
that the geometric properties of the complete physical 
evolution need not be identical to those of its 
restriction to the computational subspace. These 
observations highlight the complementary roles of 
the geometric-phase sum rule and the parallel-transport 
condition: the former constrains the total geometric 
phase accumulated by a complete basis of cyclic states, 
whereas the latter ensures the absence of dynamical 
contributions throughout the evolution.

Taken together, these results establish a fundamental 
distinction between the concatenation of Abelian 
geometric phase gates and non-adiabatic holonomic 
gates. In the former, the geometric character of 
a composite operation depends on the dynamical 
phases accumulated by its cyclic states, whereas 
in the latter, it follows from the state-independent 
parallel transport of the computational subspace 
under the ideal pulse conditions. More generally, 
our findings show that the geometric character of 
a quantum operation cannot, in general, be inferred 
solely from the properties of its constituent gates, 
but must be assessed in the context of the complete 
evolution. This observation is particularly relevant 
to the design of geometric quantum circuits: 
preserving geometric or holonomic character under 
concatenation provides a foundation for investigating 
the extent to which the associated insensitivity to 
certain perturbations can be retained at the circuit 
level, while combining these constructions with 
complementary error-resilient techniques may offer 
further improvements in robustness. Our results 
therefore establish useful criteria for designing 
universal quantum circuits based on geometric and 
holonomic principles.

\section*{Acknowledgements}
E. S. acknowledges financial support from the Swedish 
Research Council (VR) through Grant No. 2025-05249.


\begin{thebibliography}{99}
\bibitem{remark1} Experimentally, the two pulses 
are implemented over the intervals $0\leq t \leq 
\frac{\pi}{\omega}$ and $\tau \leq t \leq \tau + 
\frac{\pi}{\omega}$, for some $\tau > \frac{\pi}
{\omega}$. The pulses are turned off during the 
intervening interval $\frac{\pi}{\omega} < t < \tau $. 
In the following, we shift the time origin of the 
second pulse so that it begins at $t=0$, thereby 
setting $\tau = 0$ for notational simplicity.
\end{thebibliography}
\end{document}